# Mechanisms of Interparticle Bridging in Sintering of Dispersed Nanoparticles

Vyacheslav Gorshkov,[a] **Vasily Kuzmenko**,[a] and **Vladimir Privman**[b,*]

[a] National Technical University of Ukraine — KPI, 37 Peremogy Avenue, Building 7, Kiev 03056, Ukraine
[b] Center for Advanced Materials Processing, Department of Physics, Clarkson University, Potsdam, NY 13699, USA

**Abstract**

We model within the kinetic Monte Carlo method the initiation of neck formation and then later evolution of the resulting bridging regions for configurations involving small particles initially positioned fitted between large particles for situations typical for sintering of FCC nanocrystals, e.g., noble-metal nanoparticles. Neck initiation mechanisms by layering or clustering are identified. The stability of the resulting bridging configurations depends on several parameters, notably, on the relative small to large particle size ratio, and we explain recent experimental findings on improved sintering achieved for certain bimodal size distributions.

___

[*]Corresponding author: e-mail privman@clarkson.edu; phone +1-315-268-3891

# 1. Introduction

Sintering is an important technological process that has drawn an extensive experimental research effort.[1-9] Modeling of sintering has also been attempted, by different approaches[10-23] on the scales from continuum theories down to mesoscopic statistical mechanics descriptions, reflecting the fact that sintering is a multiscale process.[13,16,17] Specific methodologies have included continuum/finite-element,[12,14,15] atomistic kinetic Monte Carlo,[11,16,17,23] and molecular dynamics.[18] Recent use of *nanoparticles* was shown to improve the conductance[4] of sintered layers that include noble-metal nanocrystals. The particles are dispersed[2-6,8,24-28] in pastes with other additives. An interesting finding[4] that provides the motivation for the present study has been that connectivity of the metal in the resulting films can be improved by using a bimodal size distribution with smaller particles presumably fitting[4] between larger particles to facilitate bridging. Bimodal or other nonuniform size (and shape) distributions and different mixing of the particles have been considered[4,12,19,21,23,24,29-31] for potentially improving several properties of sintered materials, e.g., mechanical, density, conductance.

Here we will use the recently introduced kinetic Monte Carlo (MC) approach[23] that offers a mesoscopic-scale description of neck formation in various particle configurations. The necks can initiate bridging leading to sintering (merger) of the nearby particles, or they can be dissolved, depending on the geometry and other conditions. This process has been studied for SC lattice structures,[23] and two different nanoscale mechanisms (via clustering or layering) of neck initiation were identified. In the present study we assume FCC crystal symmetry, appropriate for noble metals. We report large scale MC simulations that elucidate the clustering and layering neck-initiation mechanisms in this case.

More importantly, we then demonstrate that the large-time fate of a neck formed due to a small particle initially positioned between two larger particles depends on the relative size of the small particle. For nanoparticles, we find that if this particle, while smaller than the particles surrounding it, is large enough then it will initiate bridging. For somewhat smaller sizes the neck that it initiates will dissolve. However, for even smaller sizes, bridging will be initiated. Really tiny particles will evaporate without any neck initiation. This is the main finding of our work because for FCC nanocrystals our estimates for the relative size of the smaller particles to initiate bridging are approximately 1:10 as compared to the larger particles, which allows the former to fit in the voids between approximately uniform larger particles in even the densest three-dimensional configurations. Thus, our finding confirms and explains the recent experimental results[4] on improved sintering that was obtained with size rations of approximately 1:7.

The dynamics of sintering[21,23,25-35] involves nonequilibrium transport of atoms, ions or molecules. We will refer to these as "atoms" for convenience. In Sec. 2, we describe the MC modeling approach to surface restructuring, atom detachment/reattachment, and diffusion in the medium. This MC approach was recently used to describe emergence of nanoparticles of well-defined shapes in synthesis,[36] and formation surface structures of interest in catalysis.[37,38]

Theoretical approaches to sintering[35] produce at best qualitative results, because sintering requires[11,12,16,17] a full mutiscale description that is presently not available. Various reported models[10-23,32-35,38-41] consider specific scales. The present mesoscopic MC approach is suitable[23,36] for the kinetics which also yields well-defined nanoparticle shapes[36,42] under appropriate synthesis conditions. The microscopic parameters enter via temperature-dependent Boltzmann weights, see Sec. 2. Then in Sec. 3 we address the mechanisms and time scales of neck formation, and in Sec. 4 we describe the results for the bridging mediated by a smaller particle located between larger particles. Both in Sec. 3 and 4 we offer preliminary explorations of the interconnections of our mesoscopic results with continuum diffusional and surface-curvature effects, as the first step towards a multiscale description. A short summary concludes Sec. 4.

## 2. Outline of the Kinetic Monte Carlo Method

The kinetic MC approach utilized here was developed for nanocrystal synthesis, and later applied to surface synthesis, and to aspects of sintering.[23,36-38,42] Therefore, we only outline its main features. During sintering, atoms can detach from nanocrystals, diffuse in the medium, and reattach. They can also locally hop on nanoparticle surfaces. In the initial configurations, such as those shown in Fig. 1, the FCC nanocrystals are assumed defectless,[36] "registered" with the underlying FCC space-covering lattice (coordination number 12) of cubic lattice spacing $a$. The detached atoms diffuse off-lattice by hopping at random angles in steps of $a/\sqrt{2}$.

Detached atoms can be reattached at vacant lattice sites nearest-neighbor to the particles. This occurs once they hop into a Wigner-Seitz cell at such locations. Atoms attached in particles can move on the surface or detach. Maintaining the precise "registration" of the attached atoms with the lattice[23,36-38] ensures that we are considering nanocrystal morphologies of relevance for particles synthesized be fast nonequilibrium techniques.[43] Such particles do not have structure-spanning defects that can control shape variation by favoring the growth of certain crystalline faces, thus resulting in unequal-proportion shapes. Nanocrystal shapes of relevance here, cf. experimental work,[4] are approximately equal-proportion, "isomeric." In reality large defects are avoided/not nucleated at the microscopic dynamics scales. However, for mesoscopic modeling our "exact registration" ensures the same result.[23,36-38]

In the initial configuration, such as in Fig. 1, each larger particle was enclosed in a rectangular box at a distance of $7a$ from the extremal coordinates of its cells. For smaller particles the box was at a distance of $5a$. The union of these boxes provided the enclosure, with reflecting boundary conditions, for the diffusing detached atoms. This mimics the fact that our few-particle configurations will in reality be part of a larger system, and we thus ensure that the detached atoms' density is controlled by their exchange with the nanocrystals and they do not dissipate to infinity. Initial particle shapes were taken those typically separately synthesized[43] (prior to their use to prepare the initial system for sintering) via fast nonequilibrium kinetics driven by plentiful supply of matter. The shapes of "isomeric" (even-proportioned) particles are then bound[23,42] by lattice planes of symmetries similar to those in the equilibrium Wulff constructions,[44-46] but with different proportions.

Each attached atom that is not fully blocked, can hop to vacant nearest neighbors. The probabilities for such moves are proportional to temperature-dependent Boltzmann factors. Each unit MC time step constitutes a sweep through the system whereby detached atoms are moved once of average, and attached atoms have on average one hopping attempt. Hoppable atoms have coordination numbers $m_0 = 1, ..., 11$. The probability for them to move is $p^{m_0}$, corresponding to a free-energy barrier, $m_0\Delta > 0$, were $p \propto e^{-\Delta/kT} < 1$. If the move it actually carried out, the atom will be repositioned to one of its $12 - m_0$ vacant neighbor sites with the probability proportional to $e^{m_f|\varepsilon|/kT}$ (normalized over all the targets). Here $\varepsilon < 0$ is the free-energy measuring binding at the target sites. The target site's coordination is $m_f = 1, ..., 11$ for hopping, and $m_f = 0$ for detachment.

Physically, we expect that connected atoms' mobility is related to the surface diffusion coefficient which is proportional to $p$, set by the free-energy scale $\Delta$, such that

$$p \propto e^{-\Delta/kT}. \tag{1}$$

Another free-energy scale, $\varepsilon$, reflects local binding, and we use its magnitude scaled per $kT$ as follows,

$$\alpha = |\varepsilon|/kT. \tag{2}$$

Results[23,36-38] on particle synthesis and surface properties, etc., suggest that within the present setting typical nonequilibrium particle morphologies are qualitatively maintained if we assume reference values $\alpha_0 = 1$ and $p_0 = 0.7$, and then temperature can be increased or decreased by changing $\alpha$, with $p$ varied according to

$$p = (p_0)^{\alpha/\alpha_0}. \tag{3}$$

For sintering, the temperature should be somewhat elevated as compared to synthesis, and therefore the present studies were with typical values of $\alpha$ in the range of 0.7 to 0.9. We actually explored many configurations and parameter values, but only representative results are summarized here.

We comment that the present simulation is large scale, carried out on clusters of 20 to 30 cores running in parallel, with CPUs such as Intel® Core™ i7-870, 2.93 GHz, or Intel® Xeon® X5660, 2.80 GHz. Various runs required from approximately 1, up to 6 weeks of CPU time, depending on the initial particle configuration, the value of $\alpha$, and the duration of the process of interest in terms of the required MC time steps.

The utilized MC approach was validated in earlier reported studies[36] for the various nanocrystal shape emergence in synthesis under dynamical conditions similar to those assumed here. Indeed, this approach could reproduce all the experimentally observed[47] metal nanocrystal shapes reported (the experiment and the relevant modeling results were for BCC). Note that in particle synthesis the atoms are supplied by external transport of matter. In sintering, considered here, the matter is actually conserved, even though atoms are constantly exchanged between the various particles and between each particle and the "gas" of diffusing single (detached) atoms. The latter was found quite dilute in our simulations reported below, with no more than a fraction of 1% of all the initially present atoms released into the diffuser "gas" at various stages of the process.

## 3. Mechanisms of Neck Initiation in Nanoparticle Sintering

In earlier work for the SC lattice symmetry, we identified[23] the cluster-formation and the layer-formation mechanisms of neck initiation between closely separated (but not in contact) nanoparticle faces. Here we explore similar neck initiation mechanisms for FCC, and we describe new findings specific for this lattice symmetry. The typical faces[36] for nanoparticle shapes for FCC in nonequilibrium synthesis for "isomeric" clusters are (111) and (100). Fig. 1 depicts (111) and (100) near-contact configurations which will be used for illustrating the differences between these two cases.

Let us first comment on the energetics of atom dynamics. We note that (111) faces are denser packed than (100) faces. Detachment of an atom from a filled (111) face requires energy $9|\varepsilon|$, whereas for (100) the energy is lower, $8|\varepsilon|$. The energies of attachment of atoms on top of fully filled faces are $3|\varepsilon|$ and $4|\varepsilon|$, respectively, which reflects the number of nearest neighbors. Therefore, surface diffusion of atoms on (111) should be faster. On the other hand, at the facet edges the detachment energies vary from $|\varepsilon|$ to $5|\varepsilon|$ for (111), but only from $|\varepsilon|$ to $3|\varepsilon|$ for (100). As a result, on average a cluster formed by absorbed atoms is more stable for (111).

Our conclusions for the neck initiation kinetics summarized in this section were obtained for various combinations of particles sizes (average radii) ranging from $20a$ to $50a$ for the large particles (the outer particles in Fig. 1), whereas the smaller particles (the middle particles in Fig. 1) had average radii from $15a$ to $30a$. Initial interparticle gaps sizes ranged up to approximately $6a$, fitting 9 lattice layers of the (111) type. (The case of a smaller middle particle, cf. Fig. 1, will be discussed in the next section.) Several realizations (at least 10 MC runs) were carried out for varying combinations of these parameters to make sure that we incorporate the possible statistical variations in our conclusions. Only selected illustrative configurations are graphically presented here.

Neck formation for facing (111) surfaces, Fig. 1(a), occurs via the layering mechanisms. This is illustrated in Fig. 2. This mechanism was studied in detail for the SC symmetry.[23] In this scenario relatively well-packed layers form consecutively in the gap, which is thus reduced to 2–3 vacant lattice layers. The final connection is established by randomly formed and diffusing on the facing surfaces few-particle clusters coming in contact and initiating rapid completion of the well-developed neck, as illustrated in the change between the last two snapshots in Fig. 2. It is interesting to note that for particles of different sizes the formation of the new layers in the gap preferentially occur on the smaller particle face. Considering several configurations, we conclude that for typical proportions and sizes as in Fig. 2, the asymmetry in the new layer formation is approximately 4-fold in favor of layering on the smaller particle. The time of the formation of a well-defined neck fluctuates from $t_{\text{neck}} = 9 \times 10^5$ to $12 \times 10^5$ MC time steps.

The observed asymmetry has interesting implications for connecting our mesoscopic-scale description to large-scale continuum modeling of aspects of sintering. Exchange of atoms between the (111) faces via the detached atom "gas" cannot on its own account for the observed formation of order 4-5 filled layers (seen in Fig. 2 from $t = 6 \times 10^5$, up to $t = 9 \times 10^5$, right before bridging) on top of the smaller of the two nanosize surfaces. Indeed, at this stage of the process we found that there is no significant loss of mass into the "gas." This is controlled by the saturation of the "gas" due to the outer boxes that mimic the rest of the sintering system, mentioned in Sec. 1. However, transport via the detachment/reattachment process and on-surface hoping leads to flow of matter on the particle surfaces leading to their rounding (illustrated in earlier work[23] for SC) and to some transfer of materials between faces of different symmetry. Specifically, here matter originating from the nearby (100) faces that have somewhat lower detachment energy than (111), is transported to (111) faces. However, considering the kinetics of the (111) surfaces that are not facing each other, we noted that at most two added layers can be generated on the relevant time scales.

The fact that the net flow of matter into the gap results in layers preferentially forming on the smaller particle that can be viewed as having on average a larger mean curvature, and that the growth is driven by that the added layers effectively increase this curvature, is a feature that can be studied within the continuum diffusion process to an absorbing surface, similar to other diffusion-limited surface growth problems.[48-50] This is outside the scope of the present work. Figure 3 highlights some of the above expectation. It provides statistics on the added layer emergence and composition that confirms that all such layers contain a significant fraction of atoms not originating from the particle on top of which they are formed. This offers evidence of a dominant role of the diffusional transport via the detached-atom "gas" in the dynamics.

The processes involved in sintering are all random, subject to thermal-noise statistical-mechanical fluctuations, modelled here within the MC approach. Specifically, we note that the formation of the neck by the layering mechanism proceeds in two stages. Once several layers are formed, to reduce the residual gap down to at most 3 missing layers for bridging, then the remaining layers are completed with rather small cross-sections and then rapidly expand to a stable neck. A snapshot for a configuration with the just-formed bridging layers is shown in Fig. 4. It illustrates the statistical aspects of the dynamics. Of particular interest are the fluctuations during the last step of the neck formation, when the connecting part broadens, because this step contributes most of the statistical noise resulting in a rather large spread of the neck initiation times, $t_{neck}$. For the realizations with parameters used for Fig. 2, 3 and 4, we already reported the range $t_{neck} = (9-12) \times 10^5$ MC time steps. For higher temperatures this time is shorter, for example, we find that $t_{neck} = (4-5) \times 10^5$ when $\alpha$ is reduced from 0.8 to 0.7. For smaller initial gap, the neck formation is also faster. For example, for a realization with all the dimensions reduced approximately 1.5-fold, with the large/small particle sizes and the gaps reduced to 35$a$, 21$a$, and approximately 4$a$ (fitting 6 lattice layers), respectively, we find that $t_{\text{neck}} = (1.8$–$3.2) \times 10^5$.

We next consider neck formation for (100) type surfaces facing each other, as in Fig. 1(b). Let us first point out that for any initial orientation of the particles of any two nanosizes, the gap must be small enough to have a large probability of neck formation. Otherwise, for larger than certain gap sizes, usually for few-particle configurations the large particles will consume matter causing small particle to dissolve and ultimately disappear. This is a well-known process of Ostwald ripening.[51] An interesting observation is offered by the case presented in Fig. 5. Here the configuration of Fig. 1(b) is considered for particle sizes and gaps similar to those used for results shown in Figs. 2 and 3, specifically, gap size 6$a$. However, in this case each gap fits 11 layers of the (100) type lattice planes, which are less dense than the (111) planes. While few added layers form on top of the large particle, see Fig. 5, with additional clustering in few extra layers, there is only limited initial clustering in a couple of layers on top of the small particle, which begins to dissolve and actually fully loses some of its original outer layers for the largest times shown. For geometrically equal conditions, this difference can be loosely attributed to that there is no flux of matter into the (100) gap form the nearby faces.

Figure 6 offers an interesting example of neck formation for the (100) facing nanosize surfaces in the geometry with sizes reduced to have the gap of only 4a, fitting here 7 lattice layers, as compared to 6 layers for the case of (111) facing surfaces, the neck-formation time range for which was reported earlier. We observe that the outer fluctuations reaching into the gap by partially filling layers on top of the particles (which do not dissolve) are clusters rather than largely filled layers. Once the initial contact is established, here also the neck is formed and fills up rapidly. As long as this mechanism of bridging is active (instead of the smaller particle dissolving), the neck formation by tenuous clusters is actually faster than by layering. For this geometry, neck formation is approximately 3 times faster for (100) facing surfaces than for (111).

The nature of the clustering growth process on nanosize faces was discussed in detail for the SC lattice symmerty.[23] For nanosize surfaces, energy considerations mentioned earlier suggest that atoms for (100) can be easier transported off the layer edges into the next layer without losing contact with the underlying structure than for the (111) case, whereas surface diffusion that allows expansion of continuous layers is faster for (111). Only a couple of nearly filled layers can form on top of the original nanosize faces before the growth reverts to the cluster node for (100). Since clusters cannot reach too far, it transpires that approximately 7 lattice layers is the maximum gap for which the clustering mechanism will lead to a likely neck formation. For larger gaps, the small particle dissolves. Clustering mechanism works for smaller initial gaps with the formation time noticeably decreasing. For example, lowering the gap by just a single layer, to 6 rather than 7 lattice layers, with all the other parameters the same as in Fig. 6, approximately halves the mean neck formation time, to $\langle t_{neck} \rangle_{average} = 0.4 \times 10^5$.

## 4. Stability of Interparticle Bridging After the Neck Initiation

Noble-metal nanoparticles are sintered in industrial processes that involve complex steps. The original particle synthesis results in a certain size and shape distribution, and also leaves organic residues at particle surfaces. They then are introduced into a paste-like "ink" to be spread on a substrate. The paste has other fillers, and its application also involves mechanical compression and contraction. Sintering at elevated temperatures is accompanied by "firing" which burns away some of the organic fillers.[25-28] Experimental data[25] on nanoparticle densities during such processing suggest that an assumption of initial gaps of order 3 to 5 atomic layers is more realistic than direct contact, largely due to the presence of organic fillers.

Particle sizes and shapes can also vary, but they are never fully uniform, and one of the interesting questions to address is the degree of nonuniformity and the shape of their size distribution for optimal products of sintering. Our present simulations for FCC and also earlier studies[23] for SC provide information on aspects of the selection of size distribution for sintering of approximately even-shapes (isomeric) nanoparticles. In Sec. 2 and in earlier studies[23] we observed that smaller particles not only form necks with larger particles but can also be dissolved into them. The latter process can also occur even after the initiation of well-formed necks.[23] Interestingly, we can qualitatively connect the consideration of this effect to another continuum-description concept, namely, the relation of the transport of matter to the mean curvatures of the sintered-structure surfaces.

Indeed, with the initial bridging to larger particles established, smaller particles' matter is pulled to its neighbors and its shape will typically be distorted from approximately even-shaped to elongated, approximately cylindrical. Geometrically, when a sphere of diameter $D$ is converted to a cylinder of height $D$ keeping the same volume (with the respective reduction of the base diameter to $\sqrt{2/3}D$), the surface curvature is changed from the sphere's $2/D$ to the cylinder's $\sqrt{3/2}/D$. The loss of matter via the detached-atom gas is typically considered linearly related to the curvature.[51] Similarly, on-surface transport should also be faster on more curved surfaces, although it is not clear whether its rates are linearly related to the curvature. Therefore, a ballpoint estimate of the range of radii for which a small particle will not dissolve can be obtained from the condition

$$\frac{\sqrt{3/2}}{D_{\text{small}}} \lesssim \frac{2}{D_{\text{large}}}, \tag{4}$$

or

$$D_{\text{small}} \gtrsim 0.6 D_{\text{large}}, \tag{5}$$

with of course $D_{\text{small}} < D_{\text{large}}$. We conclude that the spread of the particle size distribution should not exceed roughly 40%. This is indeed what was observed in numerical modeling for selected relatively narrow single-peaked particle size distributions.[23,52]

The above consideration relies on equilibrium concepts such as those related to the Young-Laplace relation. They however do not account for the effects of possible fast dynamical processes. Specifically, if the small particle is significantly smaller that the surrounding large particles, and furthermore the latter are rather closely positioned in a randomly packed arrangement, then the flow of matter from the small particle after the initial bridging can be fast enough to distort its shape past flat-cylindrical to one with a bottleneck in the middle. Since one of the principal curvatures at the narrowed (bottleneck) section in the middle of the bridging region is negative, the mean curvature is at some point no longer increasing and therefore the process of dissolution can slow down. As a result, the bridging region can sustain the full merger (sintering) of the large particles.

Figure 7 illustrates such a process for the initial configuration of Fig. 1(a), i.e.., for (111) facing surfaces. The middle cross-section of a bridging region formed form a rather small particle snuggly fit between two large particles is initially narrow, but later it begins to expand, as depicted in Fig. 7(c). However, Fig. 7 also illustrates the randomness of the process. The bridging region dissolved in approximately 10% of MC realizations. Furthermore, the observed effect of stable connection is very sensitive to the proximity of the particles. Recent experimental work[4] reported that improved quality of conducting sintered films was obtained when largely uniform-shaped large particles were mixed with small particles approximately 1/7 in size. This allowed compact packing of the initial configuration. Our observations explain the experimental findings.

A ballpoint estimate from our numerical studies is that the particle size ratio as small as approximately 1:10, e.g., Fig. 7, is typically required for stable bridging by this mechanism. One can also offer rather complicated geometrical considerations, not detailed here, that, similar to the arguments leading to Eq. (5), yield the criterion that a small particle can mediate such bridging when it is practically in contact and is less than the fraction $\left(\frac{2}{\sqrt{3}}\right) - 1 \simeq 15\%$ of the large particle sizes, i.e.,

$$D_{\text{large}} \gtrsim 6.5 D_{\text{small}} .\qquad(6)$$

However, in reality such arguments are not only approximate but also strongly geometry-dependent (groups of particles in a random mixture will not be initially aligned along a single direction in the considered configuration). Therefore, the ballpoint estimates of the size ratio $D_{\text{small}} : D_{\text{large}}$ being less than 1:10, 1:7, 1:6.5, suggested by numerics, experiment, and geometrical considerations, respectively, are all de-facto the same.

In summary, in this work we considered sintering of FCC nanocrystals at mesoscopic length scales for which the utilized kinetic MC method is expected to yield qualitatively descriptive results for the stages of the dynamics involving the initiation of neck formation and development of local bridging. Later-stage transport of matter on larger scales, typically accompanied by densification and other mechanical changes, requires a continuum-modeling macroscopic description.

We confirmed the existence of the layering and clustering regimes for neck initiation between facing nanosize surfaces of particles, depending on their symmetry. We argued that the accompanying energy and transport-rates considerations relate to continuum ideas of diffusional fluxes. Continuum ideas of surface-curvature effects on the transport also relate to the studied mesoscale kinetics in the regime of later evolution of the formed bridging regions and particularly the fate of the small particles initially snuggly fit between large ones.

**FIGURES and CAPTIONS**

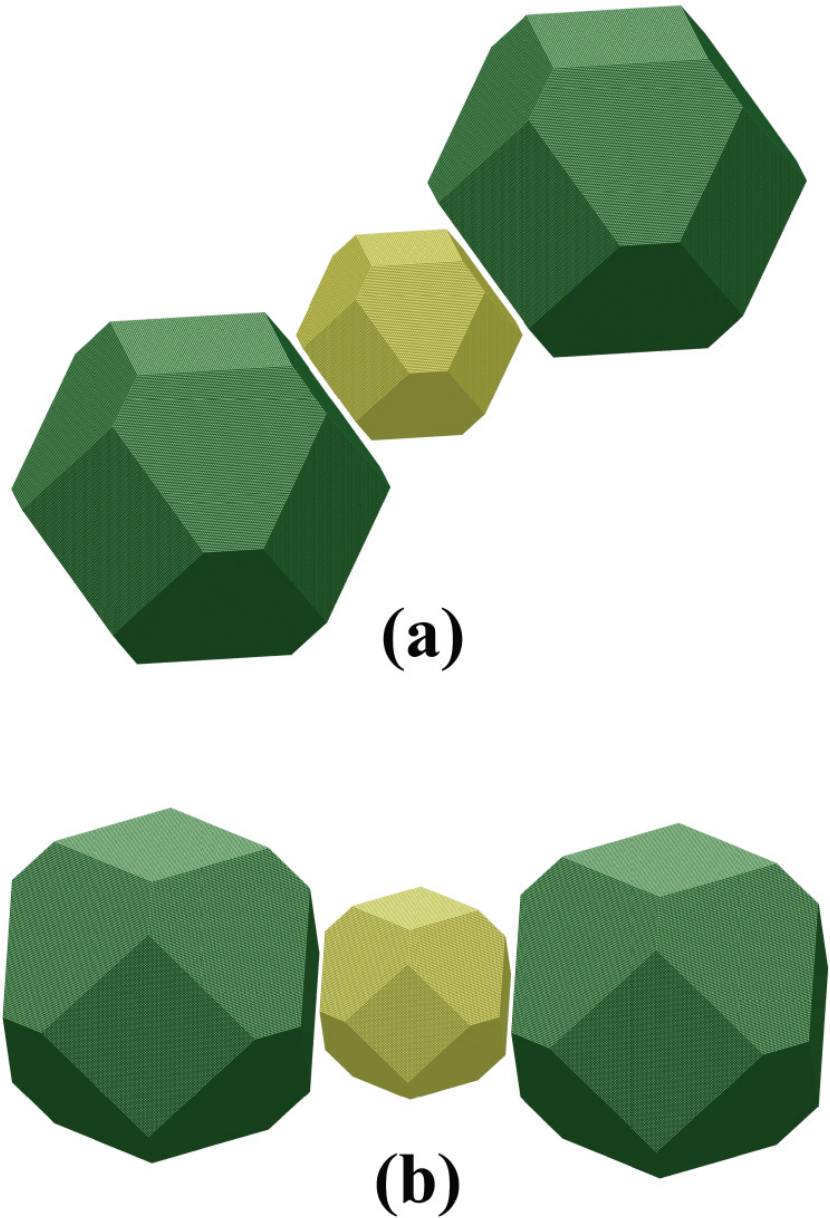

**Figure 1.** Schematic illustration of two possible initial-configuration arrangements of larger particles with a smaller particle between them, considered in the present work. (a) Here the nanocrystal surfaces facing each other are of the (111) type. (b) In this configurations, type (100) surfaces face each other.

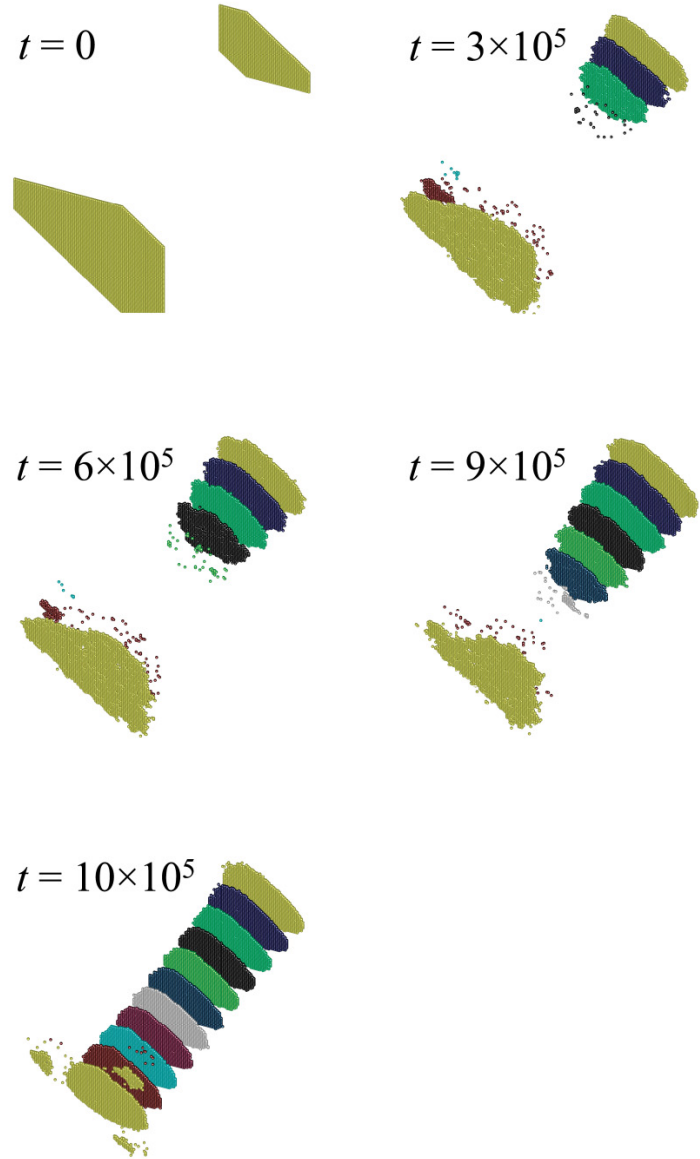

**Figure 2.** Emergence of the neck between larger and smaller particles facing each other with (111) surfaces. The initial gap size was approximately $6a$, the average particle radii were $50a$ and $30a$. The temperature parameter was $\alpha = 0.8$. The images show the surfaces of the particles and the intermediate FCC layers at $t = (0, 3, 6, 9, 10) \times 10^5$ MC time steps, as labeled in the panels. Note that the small spherical symbols represent the FCC cell centers occupied by atoms presently connected to one or both particles. The detached atoms in the vicinity are not shown here and in all the other figures. The color coding highlights the FCC lattice layers in and bordering the gap.

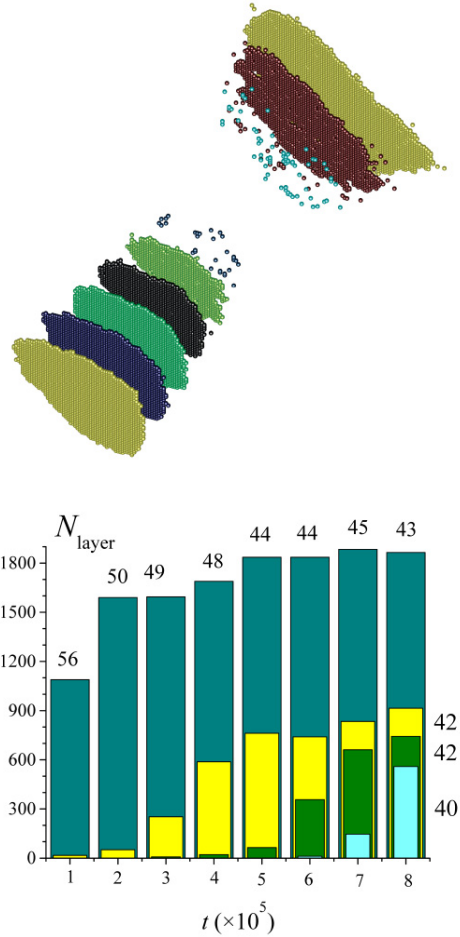

**Figure 3.** Statistics of the composition of added layers grown in a gap, on top of the small particle in a simulation with the same particle and gap sizes and the temperature parameter as in Fig. 2. The image shows the original outer particle surfaces and the added layers for a different MC run than that in Fig. 2, for time $t = 8 \times 10^5$. The bars show the time-dependence of the count of atoms in the first four added layers on top of the small particle, with the first layer, then the second layer, etc., color-coded as follows: dark cyan, yellow, olive, light cyan. The numbers added on top of the bars indicate the percentages of the atoms in the first added layer that originated from the small particle rather than were transported from the two larger particles. Such percentages for further added layers are shown only for $t = 8 \times 10^5$, on the right of the plot. Note that for this time, the only well-formed (the first) added layer on top of the large particle contained 60% of the atoms originally from that particular particle.

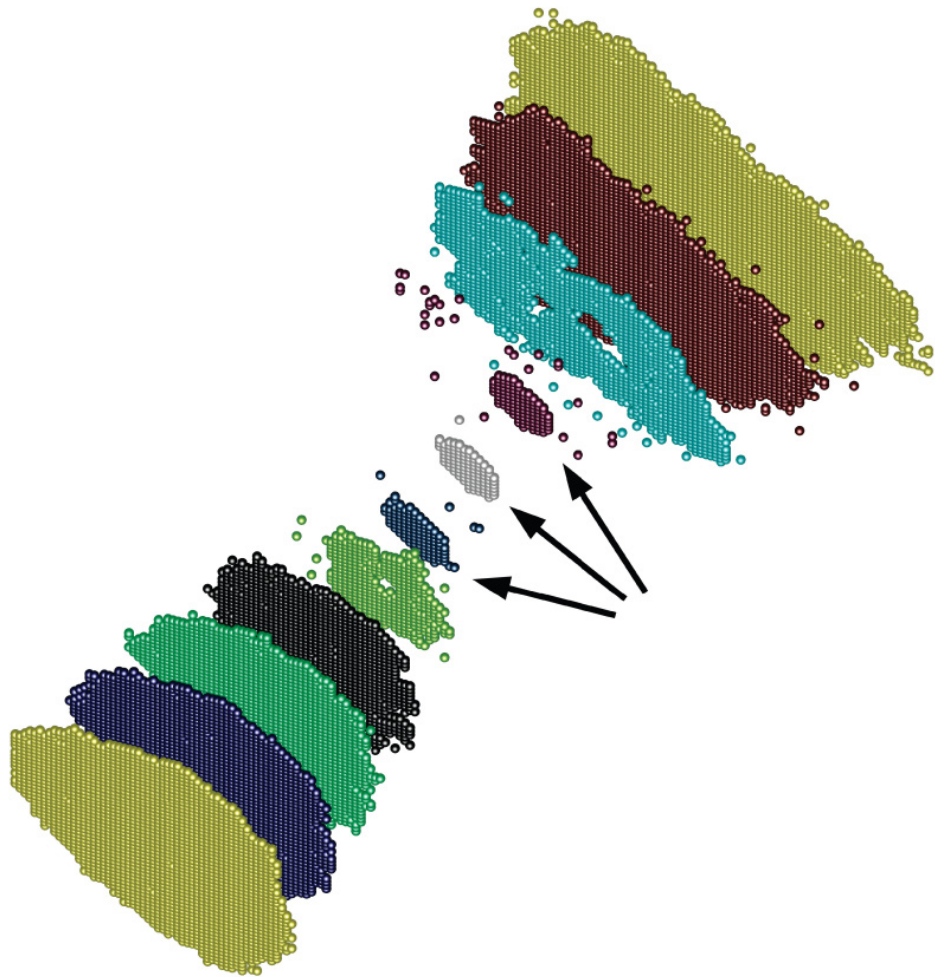

**Figure 4.** This snapshot was obtained for the same system as in Fig. 2 and 3, for the instance of time during the process that illustrates the initial stage of the formation of the last three layers (marked by arrow) providing the bridging, from which a stable neck then rapidly develops at later times. This image also illustrates the statistical nature of the dynamics, because the shown configuration was captured at $t = 10 \times 10^5$ MC steps, for which another MC realization, cf. Fig. 2, had a different neck structure. The asymmetry here is also less pronounced than for the two realizations shown in Figs. 2 and 3.

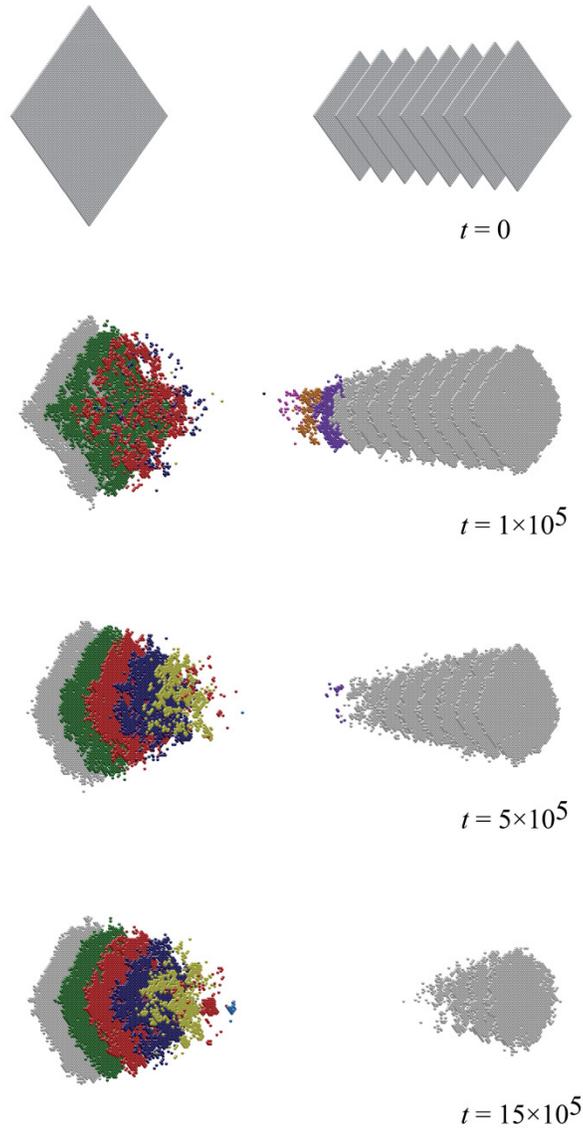

**Figure 5.** Onset of dissolution of the small particle, without neck formation, for the intial configuration of the large and small particles facing each other with (100) surfaces, Fig. 1(b). The initial average particle ($50a$ and $30a$) and gap sizes (approximately $6a$) were approximately the same as for the (111)-facing case in Figs. 2 and 3. The temperature parameter was the same, $\alpha = 0.8$. Here the 8 outer layers originally in the small particle are shown, as well as a single such layer in the large particle, color-coded gray. Other colors highlight occupied lattice sites in layers that are in the gap, for times $t = (0, 1, 5, 15) \times 10^5$ MC time steps.

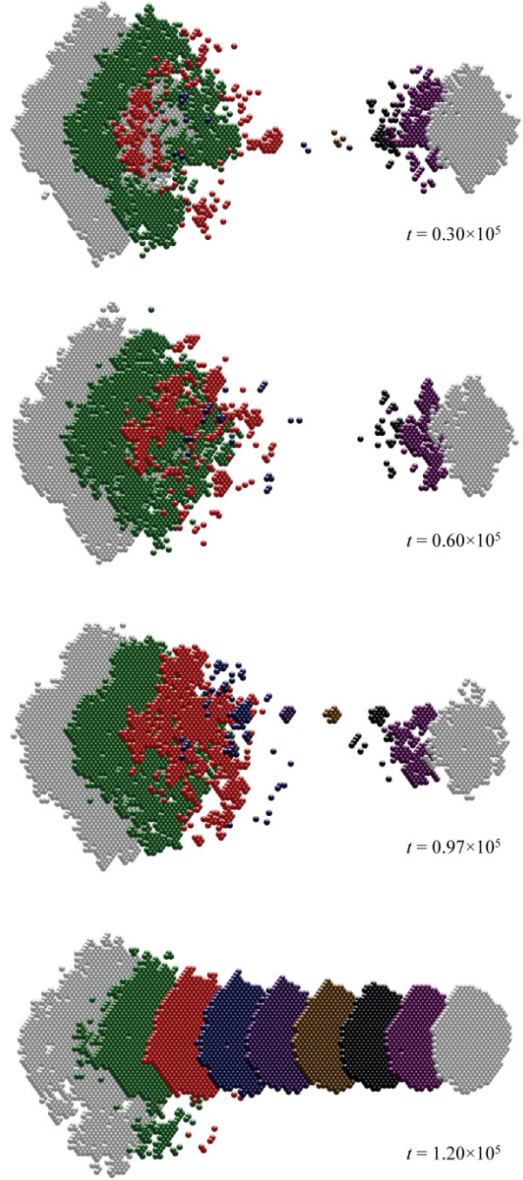

**Figure 6.** Neck development for the case of all the particle and gap sizes reduced approximately 1.5-fold as compared to Fig. 5, for the initial configuration of Fig. 1(b). Here the large/small particle sizes and the gaps were $35a$, $21a$, and approximately $4a$ (fitting 7 lattice layers), respectively. The outer particle layers are color-coded in gray, whereas other colors mark filled sites in the gap layers, for times $t = (0.30, 0.60, 0.97, 1.20) \times 10^5$. The instance of the formation of clusters that bridge the particles, followed by rapid neck emergence for this MC realization was captured at $t = 0.97 \times 10^5$.

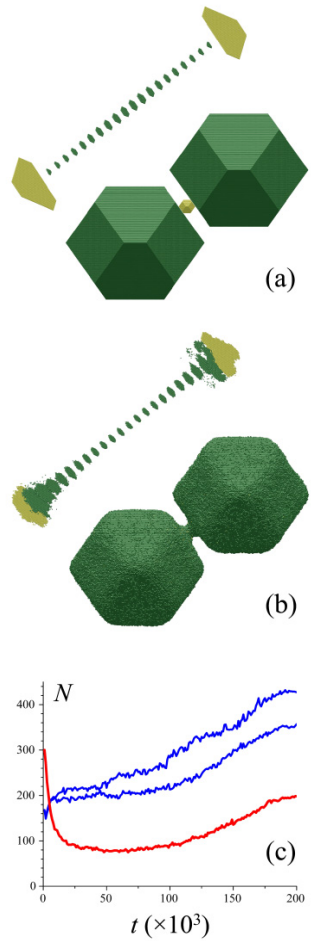

**Figure 7.** (a) The initial configuration here is that of Fig. 1(a), with the average particle radii $50a$ for both large particles, but only somewhat over $5a$ for the small particle, and with the two gaps each fitting only a single lattice layer. The interior of the region between the two large particles is shown magnified, i.e., disproportionately stretched along the axial direction. (b) The configuration at $t = 2 \times 10^5$ MC time steps. The bridged region is shown magnified (stretched) here was well. (c) Time-dependence of the total number of atoms in the middle cross-section of the connecting structure, which originated form the small particle and bridges the two large particles, is color-coded red. The blue data show time-dependence of the count of atoms half-way from the middle-cross-section to the large particles in both directions (means, in the cross-sections 1/4 and 3/4 of the way, along the axial distance between them). Besides the expected statistical noise, the feature of notice is that the middle cross section initially narrows well beyond the part-way ones, but later it actually begins to fatten.

VOLUME 2 • NUMBER 2   AUGUST 2014

www.aspbs.com/jcsmd*Journal of*

# COUPLED SYSTEMS *and* MULTISCALE DYNAMICS

*Editor-in-Chief:* Prof. Dr. Roderick Melnik, Canada

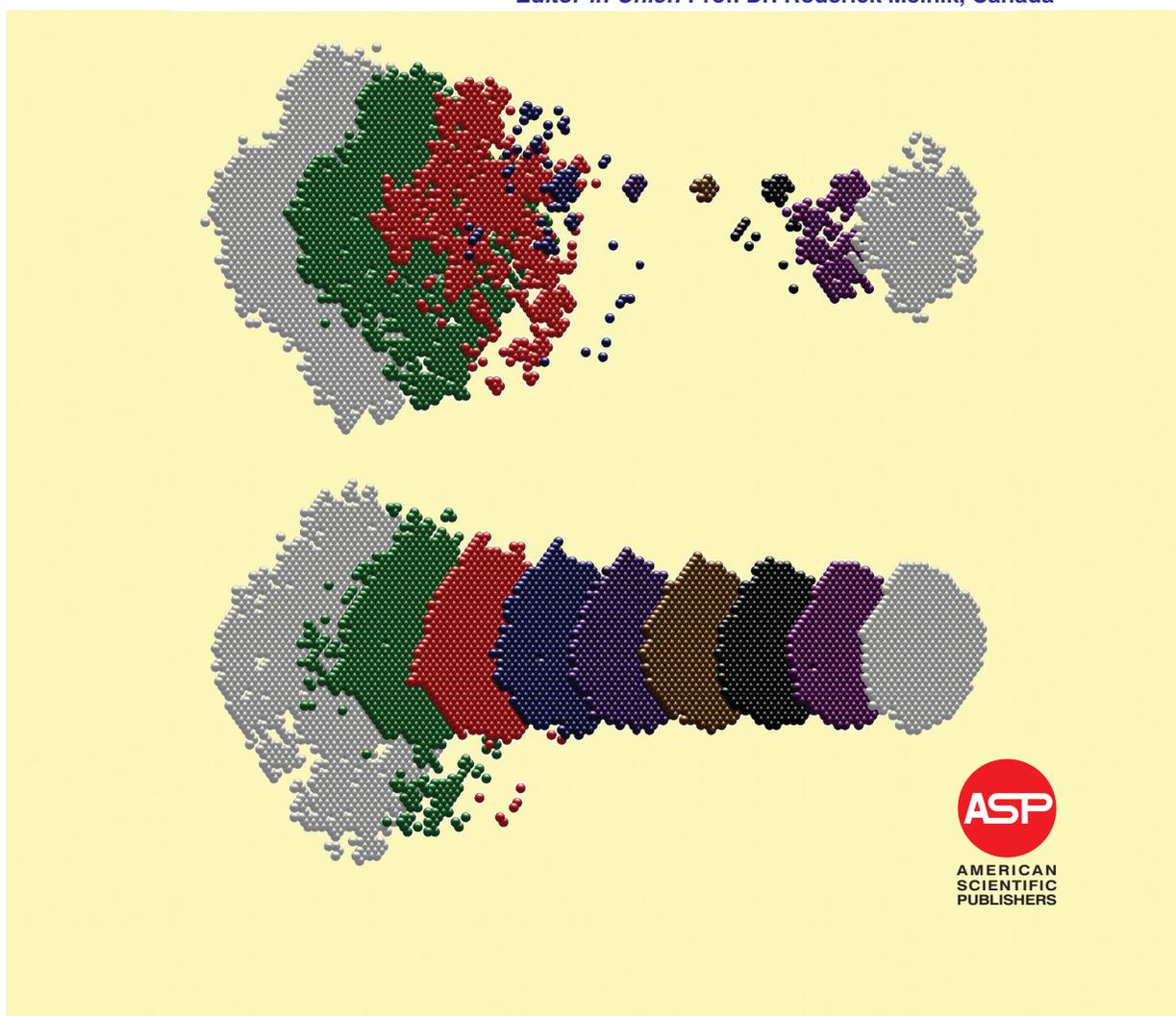

**ON THE COVER:** Initiation of neck formation is shown, modeled within the kinetic Monte Carlo approach that was also applied to the evolution of the resulting bridging regions for configurations involving small particles initially positioned fitted between large particles in sintering of noble-metal nanocrystals. Neck initiation mechanisms by layering or clustering were identified, and the stability of the resulting bridging was found to depend on several parameters including the relative particle sizes, explaining recent experimental findings on improved sintering achieved for certain bimodal size distributions. [Credits: *Vyacheslav Gorshkov, Vasily Kuzmenko, and Vladimir Privman,* "Mechanisms of Interparticle Bridging in Sintering of Dispersed Nanoparticles", J. Coupled Syst. Multiscale Dyn. 2 (2), 91–99 (2014).]